\documentclass{appolb}
\usepackage{graphicx}
\usepackage{amsmath}
\usepackage{amstext}
\usepackage{amssymb}

\begin{document}
\title{Electromagnetic probes of the Quark-Gluon Plasma%
\thanks{Presented at XXIX$^{\rm th}$ International Conference on Ultra-relativistic Nucleus-Nucleus Collisions (Quark Matter 2022).}%
}
\author{Gojko Vujanovic\thanks{Speaker}$^\mathrm{a,b}$
\address{$^\mathrm{a}$Department of Physics and Astronomy, Wayne State University, Detroit MI 48201.}
\address{$^\mathrm{b}$Department of Physics, University of Regina, Regina, SK S4S 0A2, Canada.}
\\[3mm]
}
\maketitle
\begin{abstract}
The penetrating nature of electromagnetic probes makes them an ideal candidate to study properties of the Quark-Gluon Plasma (QGP). A selection of recent developments in the theory and phenomenology of electromagnetic probes is discussed, with an emphasis given towards how these probes can be used to constrain QGP trapnsport coefficients. A Bayesian treatment of electromagnetic radiation, similar to the one of soft hadronic observables and jets, is suggested as a path towards imposing more stringent constraints on various transport coefficients of the QCD medium. 
\end{abstract}
  
\section{Introduction}
One of the main goals of high-energy heavy-ion collision experiments is to create a medium where the fundamental degrees of freedom of nuclear matter, the quarks and gluons, are exposed creating a Quark Gluon Plasma (QGP). As quarks have both a color charge and an electromagnetic charge, they can not only interact with the QGP via the strong force, but can also radiate photons. As photons themselves don't a carry an electromagnetic charge, while the electromagnetic coupling is much smaller than the strong coupling, photons seldom interact with the QGP once they produced. This allows photons to carry away precise information about the state the QGP was in when they were created. Detection of electromagnetic radiation is therefore vital in any precise study of QGP properties. 

The QCD medium has two sources of electromagnetic radiation. At high temperatures $T\gtrsim 0.16$ GeV, partonic interactions in the QGP are responsible for the majority of electromagnetic (EM) emissions. As lower temperatures are reached however, hadronic processes, be it hadron-hadron scattering or hadronic decay, are sourcing EM radiation. EM emission from hadronic interactions continues even when the nuclear medium is too dilute for a hydrodynamical descriptions to be valid. Whether late-time hadronic interactions producing EM radiation are included in the measured signal depends on experimental capabilities.

In heavy-ion experiments, two sources of EM probes have been detected: direct photons and lepton pairs (dileptons). The latter come from decay of virtual photons and thus have an additional degree of freedom, namely the center of mass energy of the lepton pair or the invariant mass $M$. At lower temperatures $T\lesssim 0.16$ GeV, virtual photon radiation comes mostly form decay of $\rho$, $\omega$, and $\phi$ vector mesons whose in-medium properties generate a diverse set of features seen in the EM spectral function at $M \lesssim 1$ GeV \cite{Vujanovic:2019yih}. At higher temperatures $T\gtrsim 0.16$ GeV, QGP dileptons are preferentially emitted at intermediate invariant masses ($1\lesssim M \lesssim 2.5$ GeV) and produce a rather featureless spectrum. This variation is spectral shape can be used to identify these dileptons sources. For $M\gtrsim 2.5$ GeV, decay of quarkonia consitutes an important dilepton signal as well as Drell-Yan processes. The former will not be discussed here, while the latter is produced before the QGP medium is created. As this contribution focuses on how EM probes can be used to study QGP properties, Drell-Yan processes will not be explored. Unlike dileptons, direct photons form the QCD medium produce a rather featureless spectrum, both from the hadronic interactions emitted at $p_T\lesssim 1$ GeV, and from partonic processes (at $1 \lesssim p_T\lesssim 4$ GeV) \cite{Paquet:2015lta}. 

The most direct way to measure the creation of the QGP in heavy-ion collisions is to detect a non-trivial dilepton anisotropic flow ($v_2$) at $1\lesssim M \lesssim 2.5$ GeV \cite{Vujanovic:2013jpa}, as no dilepton $v_2$ has been reported in proton-proton collision experiments. However, that invariant mass region has an additional source of dileptons stemming form semi-leptonic decay of open heavy flavor hadrons; a source that dominates over direct QGP radiation. Thus, to expose direct QGP radiation, a Heavy Flavor Tracker (HFT) needs to be installed --- such as the one at present in the STAR experiment at Brookhaven National Laboratory's Relativistic Heavy Ion Collider (RHIC) --- to remove the heavy flavor contribution to dileptons. The presence of the open heavy flavor/anti-flavor in the dilepton signal (especially dilepton $v_2$) also presents an opportunity to independently study how heavy flavor/anti-flavor {\it pairs} interact with QGP; a worthy scientific endeavor that goes beyond simple signal-removal and is only possible with an dedicated HFT instrument.

\section{Electromagnetic radiation form nuclear matter}
For both photon and dilepton sources, the key theoretical quantity being probed is the EM spectral function. The dilepton and photon production rates are given by:
\begin{eqnarray}
\frac{d^4 R_{\ell^+\ell^-}}{d^4 q} &=& -\frac{\alpha^2_{EM}}{\pi^3 M^2} \frac{{\rm Im}\left[\Pi_{EM}(M,{\bf q};T)\right]}{e^{ q\cdot u/T} -1}\nonumber\\
q^0 \frac{d^3 R_\gamma}{d^3 {\bf q}} &=&  -\frac{\alpha_{EM}}{\pi^2} \frac{{\rm Im}\left[\Pi_{EM}(M=0,{\bf q};T)\right]}{e^{q\cdot u/T} -1},
\label{eq:th_rates}
\end{eqnarray}
where ${\rm Im}\left[\Pi_{EM}\right]={\rm Im}\left[g_{\mu\nu} \Pi^{\mu\nu}_{EM}\right]$ is the EM spectral function, $T$ is the temperature,  $u^\mu$ is the flow velocity of the QGP, $\alpha_{EM}$ is the fine structure constant, and $M^2=\left(q^0\right)^2-\vert{\bf q}\vert^2$. In recent years, a vigorous effort has focused on characterizing the in-equilibrium EM spectral function using both perturbative and non-perturbative approaches. Furthermore, given how important viscosity is in modern hydrodynamical simulations of the QGP, an additional endeavor studied the non-equilibirium, viscous, effects on EM production rates in Eq. (\ref{eq:th_rates}), on both the hadronic and partonic rates.

\subsection{Electromagnetic radiation in thermal equilibrium}
The electromangetic spectral function has been calculated using both perturbative QCD (pQCD) and non-perturbative lattice approaches. The most recent effort on the pQCD side has yielded next-to-leading (NLO) EM spectral functions for both real ($M=0$) photons \cite{Ghiglieri:2013gia,Ghiglieri:2016tvj} and virtual photons, which subsequently decay into dileptons \cite{Laine:2013vma,Ghisoiu:2014mha,Ghiglieri:2014kma,Jackson:2019mop,Jackson:2019yao}. Ref. \cite{Jackson:2019yao} in particular shows the remarkable agreement between lattice QCD calculations of EM spectral functions, in quenched and unquenched approximations \cite{Brandt:2017vgl,Brandt:2019shg}, and NLO pQCD calculations. This agreement highlights the start of a new era of precision calculations of EM spectral functions. 

On the hadronic side, two streams of investigations have been explored. The first stream uses effective Lagrangians to describe many-body hadronic interactions, with Vector meson Dominance Model (VDM) employed to couple hadrons to photons. At tree-level, dileptons rates from vector mesons have been obtained through the forward scattering-based approach \cite{Eletsky:2001bb,Martell:2004gt,Vujanovic:2009wr} describing interactions between vector mesons and other hadrons, while SU(3) Massive Yang-Mills theory was used to compute photon production rates from mesonic interactions \cite{Turbide:2003si}. Interactions beyond tree-level, both mesonic and baryonic, have also been considered \cite{Rapp:1999ej,Rapp:2009yu} and their parametrization tabulated \cite{Turbide:2003si, Heffernan:2014mla}. Baryonic interactions, in particular, are important for describing the in-medium vector meson spectral functions \cite{Rapp:1999ej,Rapp:2009yu}, especially the $\rho$ meson. A recent study \cite{Hohler:2013eba} has shown that including baryonic and mesonic interactions modify not only the $\rho$ meson, but also its chiral partner, the $a_1$ meson, such that their spectral functions start overlapping as temperatures beyond 150 MeV. Thus Ref.~\cite{Hohler:2013eba} shows a first indication that chiral symmetry may be restored at high temperatures.

The second stream of investigation relies on the non-perturbative Functional Renormaliazation Group (FRG) approach. A recent study using mesonic degrees of freedom in the FRG approach \cite{Tripolt:2017zgc} shows that the $\rho$ and $a_1$ spectral functions overlap at $T=300$ MeV. Including baryonic degrees of freedom in that calculation may bring the overlap temperature closer to the pseudo-critical temperature $T=156 \pm 1.5$ MeV reported by lattice QCD calculations \cite{Bazavov:2018mes}. Future studies are needed to confirm whether that is the case however. 
\subsection{Non-equilibrium corrections to electromagnetic production rates}
Modern hydrodynamical calculations include the effects of bulk and shear viscosity, thereby modifying the underlying particle distribution away from its equilibrium form. The EM production rates at leading order can be described using the Boltzmann equation:
\begin{eqnarray}
\frac{d^3 R_\gamma}{d^3 {\bf q}} &=& \int \frac{d^3 p_1}{2p^0_1 (2\pi)^3} \frac{d^3 p_2}{2p^0_2 (2\pi)^3}  \frac{d^3 P}{2P^0 (2\pi)^3} \left(2\pi\right)^4 \delta^{(4)}\left(p_1+p_2-P-q\right) \times \nonumber \\
&\times& f_{\bf p_1} f_{\bf p_2} \frac{\left\vert \mathcal{M} \right\vert^2}{2q^0(2\pi)^3} \left(1\pm f_{\bf P}\right),
\end{eqnarray}
where $f_{\bf p_1}$ and $f_{\bf p_2}$ are incoming particles' momentum distribution, while $f_{\bf P}$ is the outgoing particle's momentum distribution, with $q^\mu$ labeling the photon's momentum and $\mathcal{M}$ being the tree-level scattering matrix element. Expanding around an equilibrium distribution $f_0$, photon production explored in  Refs. \cite{Dion:2011pp,Shen:2014nfa,Paquet:2015lta} uses the Chapman-Enskog/14-moment approximations as ans\" atze to write $f=f_0+\delta f$, with $\delta f$ encapsulating the effects of bulk and shear viscous corrections.

For dilepton production in the hadronic hadronic sector, the effects of viscosity on in-medium vector mesons was explored in the limit the leading density expansion \cite{Eletsky:2001bb,Martell:2004gt,Vujanovic:2009wr}, and the effects on dilepton observables presented in Refs.~\cite{Vujanovic:2013jpa,Vujanovic:2017psb,Vujanovic:2019yih}.
\section{Electromagnetic probes of high-energy nuclear collisions}
\begin{figure}[!h]
\begin{tabular}{cc}
\includegraphics[width=0.5\textwidth]{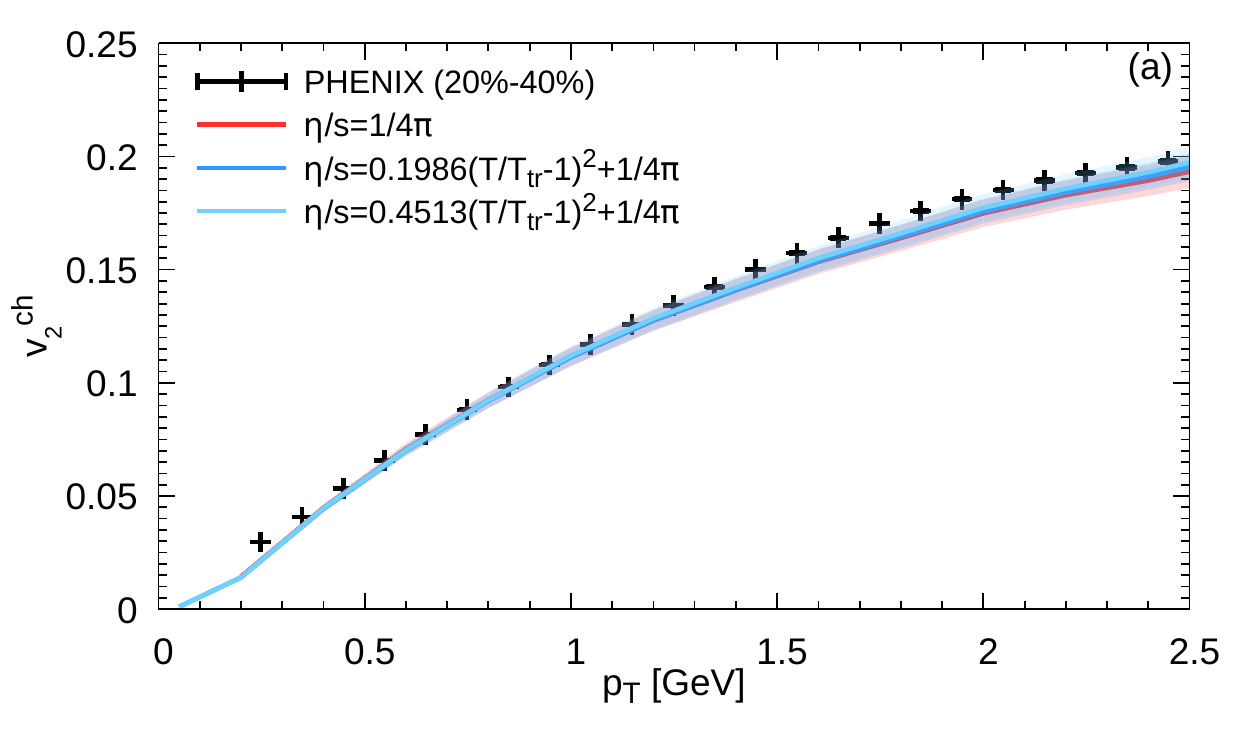} & \includegraphics[width=0.5\textwidth]{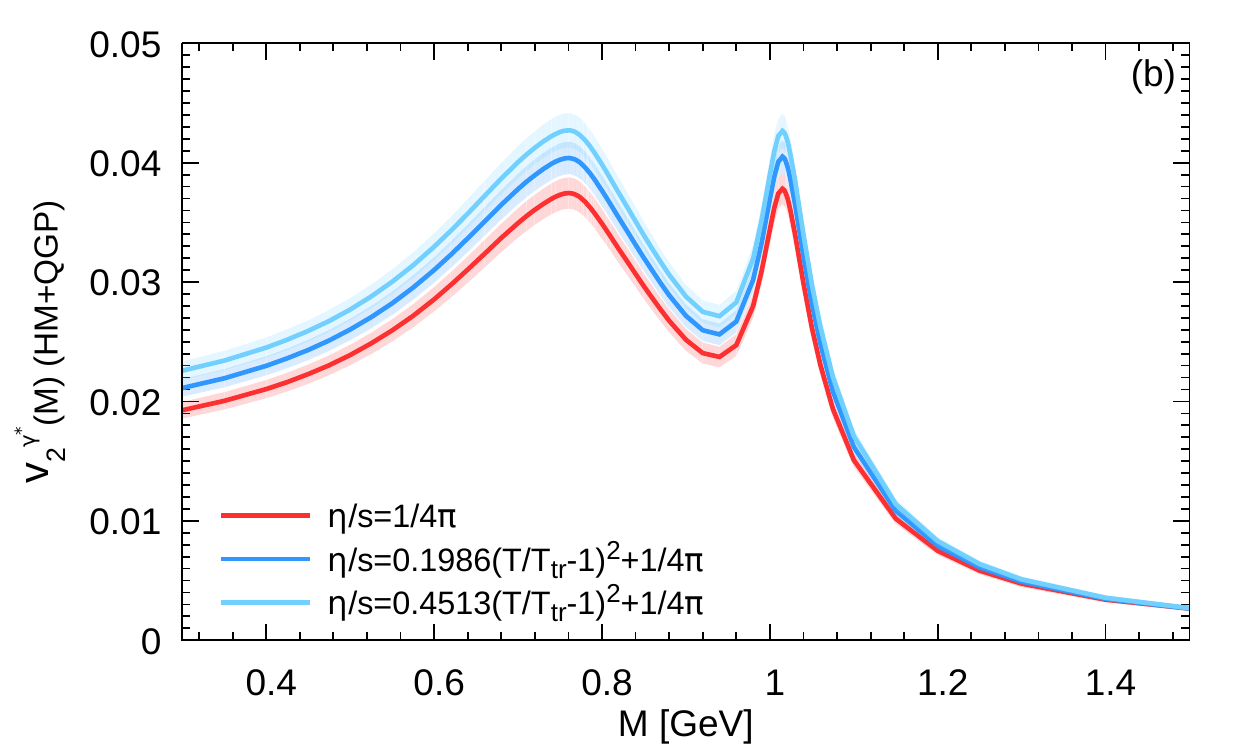} 
\end{tabular}
\caption{Charged hadron $v^{\rm ch}_2$ and dilepton $v_2(M)$ within $20-40$\% centrality at top RHIC collision energy taken from \cite{Vujanovic:2017psb}. This dilepton calculation includes sources from the hadronic sector and the QGP.}
\label{fig:v2_M_vs_ch}
\end{figure}

Electromagnetic probes are particularly sensitive to various transport coefficients of the QGP \cite{Paquet:2015lta,Vujanovic:2017psb,Vujanovic:2019yih,Vujanovic:2016anq,Kasmaei:2018oag,Kasmaei:2019ofu,Floerchinger:2021xhb}. In particular, dilepton production from a temperature-dependent $\eta/s$ has shown that they are particularly sensitive to the high-temperature behaviour of $\eta/s$, as can be seen in Fig.~\ref{fig:v2_M_vs_ch}. That result is particularly interesting given the recent Bayesian analysis' \cite{JETSCAPE:2020shq} poor constraint $\eta/s(T)$ at high temperatures. Dileptons can thus be used to further constrain specific shear viscosity at high temperatures. The constraining power of dileptons depends on the accuracy and precision with which $v_2$ is measured. Hence, high accuracy measurements are currently being planed \cite{Citron:2018lsq}. 
However, as there are additional sources of EM production beyond the hydrodynamical one, EM probes' sensitivity to QGP tranpsort coefficients may be reduced.

\begin{figure}[!h]
\includegraphics[width=0.6\textwidth]{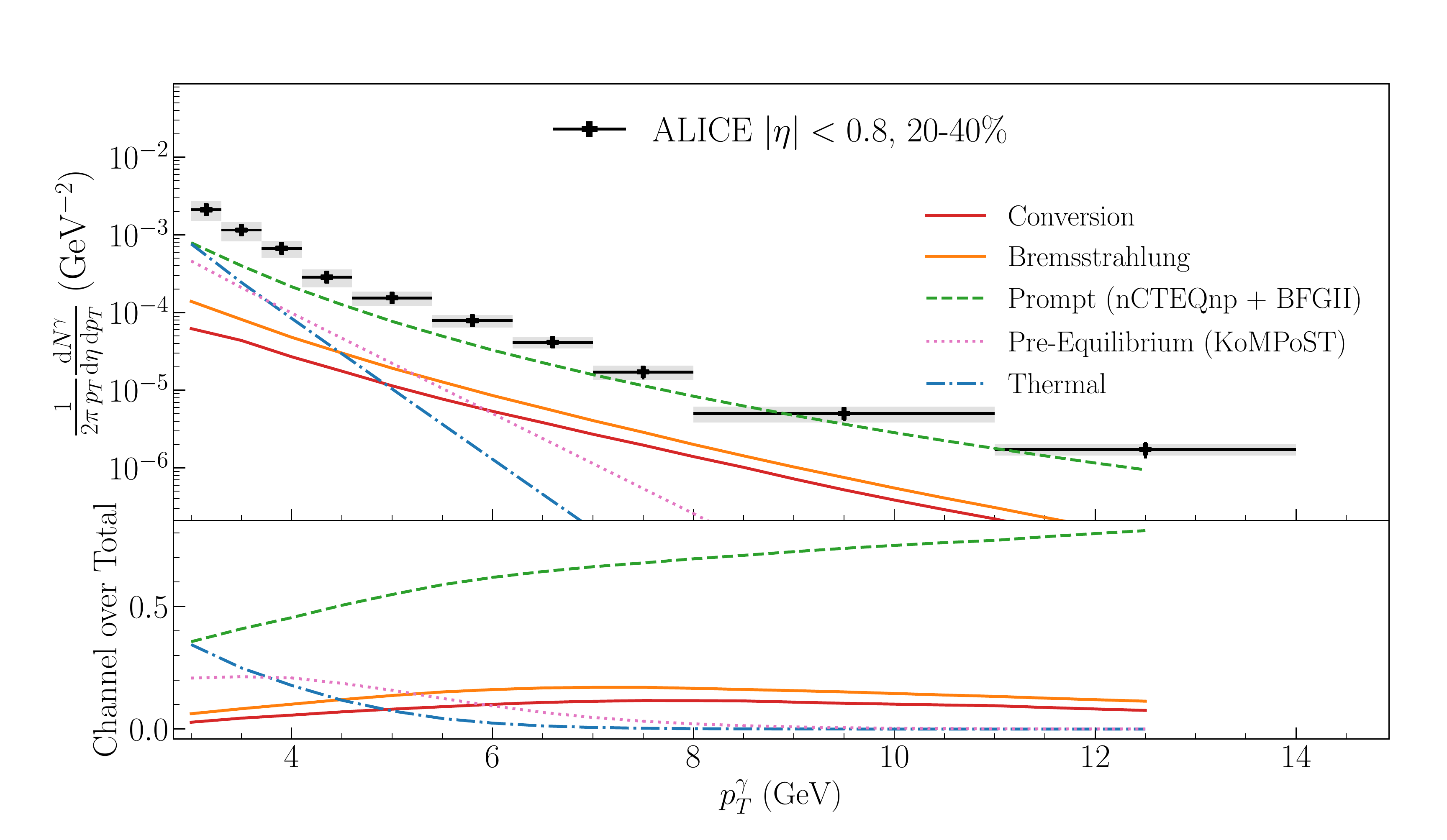}
    \begin{minipage}[b]{10pc}
    \begin{center} 
    \caption{Calculation of photon production channels in from Pb-Pb at $\sqrt{s_{NN}}=2.76$ TeV collisions at $20$-$40\%$ centrality taken from \cite{Yazdi:2022cuk}. Data from \protect~\cite{ALICE:2015xmh}.
    \label{fig:photons_by_channnel}}
    \end{center}
    \end{minipage}
\end{figure}

In the case of photon production, sources from a pre-hydrodynamical evolution \cite{Gale:2021emg} have been considered. Ref.~\cite{Gale:2021emg} shows how sensitive direct photon $v_2$ is to chemical equilibration processes taking place before the QGP hydrodynamizes. Following the QGP, photons from Boltzmann kinetic transport are highly sensitive to an off-equilibrium evolution. As a result, a recent calculation using SMASH hadronic transport \cite{Schafer:2021slz} has found large increase in photon $v_2$ from off-equilibrium dynamics. Thus, many different photons sources have to be combined in an effort to explain the discrepancy seen between photon $v_2$ calculations and experimental data, i.e. the ``photon flow puzzle''. At intermediate $p_T$, a significant portion of photon production comes from jet-medium interactions, as shown in Figure~\ref{fig:photons_by_channnel}. Indeed, Ref.~\cite{Yazdi:2022cuk} shows that $\approx 30\%$ of photon emitted within the intermediate $p_T$ range ($5\lesssim p_T \lesssim 8$ GeV) stem from jet-medium interactions.
Thus photons are not only sensitive viscosity of the QGP, but also to jet-related transport coefficients such as $\hat{q}$. This sensitivity to $\hat{q}$ is noteworthy as it avoids hadronization effects, thus making
photons a noteworthy probe for constraining $\hat{q}$. 

On the dilepton side, at intermediate invariant masses ($1\lesssim M\lesssim 2.5$ GeV), the semi-leptonic decays of open heavy (anti-)flavor pairs contribute significantly to both dilepton yield and $v_2$ \cite{Vujanovic:2013jpa}. As the heavy (anti-)quark pair traverses the QGP, the amount of energy/momentum exchange depends on both kinematics (such as the virtuality of the heavy quark), as well as on the local energy-loss properties of the QGP, such as $\hat{q}$. The evolution of the heavy quark within the QGP is best achieved using a model-agnostic framework such as that developed by the JETSCAPE Collaboration \cite{Vujanovic:2020wuk,Fan:2020afj}. Following hadronization, the dilepton signal originates from the decay of the open heavy (anti-)flavor {\it pair} and is sensitive to the variation in QGP evolution along each heavy (anti)quark's path. This is unlike other heavy flavor measurements that may not consider open heavy flavors as a pair. The heavy quark energy/momentum interaction with the QGP also generates dilepton $v_2$ \cite{Vujanovic:2013jpa}. Thus, measuring dilepton $v_2$ in the intermediate mass region is crucial to constraining $\hat{q}$ from open heavy flavor decay pairs, while removing them allows direct radiation from the QGP, giving better access to bulk properties such as viscosities. 

As $\sqrt{s_{NN}}$ is lowered, the phase space for producing jets becomes narrower and the only remaining penetrating probe to study the nuclear medium is electromagnetic. At lower collisions energies a QCD first order phase transition becomes a possibility. The consequences on a first order phase transition on dilepton production can only be fully studied in simulations that combine dileptons from a hydrodynamical simulations with those from hadronic transport \cite{Endres:2015egk,Endres:2016tkg,Staudenmaier:2017vtq,Hirayama:2022rur}. A recent calculation of dilepton production at lower collision energies \cite{Seck:2020qbx} has shown evidence that a first order phase transition can have a large effect on dilepton yield. To confirm these results, more studies are needed in the future. 
\section{Conclusion}
Though rarely produced, EM probes are simultaneously sensitive to bulk medium transport coefficients, such as $\frac{\eta}{s}(T)$, as well as jet-related transport coefficients, e.g., $\hat{q}$. For the latter, EM probes avoid hadronization effects, making them extremely valuable. The next step in phenomenological calculations of EM radiation is to combine different sources together within a Bayesian analysis aiming at improving constraints on transport coefficients. Of course, theoretical improvements, such as better $\delta f$ calculations, should be pursued in parallel as they contribute to the theoretical systematic uncertainty on QGP transport properties, and need to be included in upcoming Bayesian model-to-data comparisons. Bayesian analysis using dileptons can also investigate whether chiral symmetry restoration effects are seen in data. This can be achieved using Bayesian model selection, where a model-to-data comparisons of calculations with and without chiral symmmetry restoration can ascertain whether experimental data favors either calculation. 

Bayesian analyses could also advise whether better measurements are needed in certain areas, thus increasing the synergy between theory and experiment. To that end, a Bayesian meta-analysis including EM probes, soft hadronic observables, and jet-related measurement are need within a holistic approach to understanding QGP properties.

{\bf Acknowledgements}: This work was supported by the Natural Sciences and Engineering Research Council of Canada, and by the National Science Foundation (in the framework of the JETSCAPE Collaboration) through award No. ACI-1550300.

\bibliographystyle{h-physrev3.bst}
\bibliography{references}

\begin{thebibliography}{10}

\bibitem{Vujanovic:2019yih}
G.~Vujanovic {\em et~al.},
\newblock Phys. Rev. C {\bf 101}, 044904 (2020).

\bibitem{Paquet:2015lta}
J.-F. Paquet {\em et~al.},
\newblock Phys. Rev. {\bf C93}, 044906 (2016).

\bibitem{Vujanovic:2013jpa}
G.~Vujanovic {\em et~al.},
\newblock Phys. Rev. {\bf C89}, 034904 (2014).

\bibitem{Ghiglieri:2013gia}
J.~Ghiglieri {\em et~al.},
\newblock JHEP {\bf 05}, 010 (2013).

\bibitem{Ghiglieri:2016tvj}
J.~Ghiglieri, O.~Kaczmarek, M.~Laine, and F.~Meyer,
\newblock Phys. Rev. {\bf D94}, 016005 (2016).

\bibitem{Laine:2013vma}
M.~Laine,
\newblock JHEP {\bf 1311}, 120 (2013).

\bibitem{Ghisoiu:2014mha}
I.~Ghisoiu and M.~Laine,
\newblock JHEP {\bf 1410}, 83 (2014).

\bibitem{Ghiglieri:2014kma}
J.~Ghiglieri and G.~D. Moore,
\newblock JHEP {\bf 1412}, 029 (2014).

\bibitem{Jackson:2019mop}
G.~Jackson,
\newblock Phys. Rev. D {\bf 100}, 116019 (2019).

\bibitem{Jackson:2019yao}
G.~Jackson and M.~Laine,
\newblock JHEP {\bf 11}, 144 (2019).

\bibitem{Brandt:2017vgl}
B.~B. Brandt, A.~Francis, T.~Harris, H.~B. Meyer, and A.~Steinberg,
\newblock EPJ Web Conf. {\bf 175}, 07044 (2018).

\bibitem{Brandt:2019shg}
B.~B. Brandt {\em et~al.},
\newblock {Lattice QCD estimate of the quark-gluon plasma photon emission
  rate},
\newblock in {\em {37th International Symposium on Lattice Field Theory}},
  2019, 1912.00292.

\bibitem{Eletsky:2001bb}
V.~L. Eletsky, M.~Belkacem, P.~J. Ellis, and J.~I. Kapusta,
\newblock Phys. Rev. {\bf C64}, 035202 (2001).

\bibitem{Martell:2004gt}
A.~T. Martell and P.~J. Ellis,
\newblock Phys. Rev. {\bf C69}, 065206 (2004).

\bibitem{Vujanovic:2009wr}
G.~Vujanovic, J.~Ruppert, and C.~Gale,
\newblock Phys. Rev. {\bf C80}, 044907 (2009).

\bibitem{Turbide:2003si}
S.~Turbide, R.~Rapp, and C.~Gale,
\newblock Phys. Rev. C {\bf 69}, 014903 (2004).

\bibitem{Rapp:1999ej}
R.~Rapp and J.~Wambach,
\newblock Adv.Nucl.Phys. {\bf 25}, 1 (2000).

\bibitem{Rapp:2009yu}
R.~Rapp, J.~Wambach, and H.~van Hees,
\newblock Landolt-Bornstein {\bf 23}, 134 (2010).

\bibitem{Heffernan:2014mla}
M.~Heffernan, P.~Hohler, and R.~Rapp,
\newblock Phys. Rev. C {\bf 91}, 027902 (2015).

\bibitem{Hohler:2013eba}
P.~M. Hohler and R.~Rapp,
\newblock Phys. Lett. B {\bf 731}, 103 (2014).

\bibitem{Tripolt:2017zgc}
R.-A. Tripolt, B.-J. Schaefer, L.~von Smekal, and J.~Wambach,
\newblock Phys. Rev. D {\bf 97}, 034022 (2018).

\bibitem{Bazavov:2018mes}
HotQCD, A.~Bazavov {\em et~al.},
\newblock Phys. Lett. B {\bf 795}, 15 (2019).

\bibitem{Dion:2011pp}
M.~Dion {\em et~al.},
\newblock Phys. Rev. C {\bf 84}, 064901 (2011).

\bibitem{Shen:2014nfa}
C.~Shen, J.-F. Paquet, U.~Heinz, and C.~Gale,
\newblock Phys. Rev. {\bf C91}, 014908 (2015).

\bibitem{Vujanovic:2017psb}
G.~Vujanovic, G.~S. Denicol, M.~Luzum, S.~Jeon, and C.~Gale,
\newblock Phys. Rev. {\bf C98}, 014902 (2018).

\bibitem{Vujanovic:2016anq}
G.~Vujanovic {\em et~al.},
\newblock Phys. Rev. {\bf C94}, 014904 (2016).

\bibitem{Kasmaei:2018oag}
B.~S. Kasmaei and M.~Strickland,
\newblock Phys. Rev. D {\bf 99}, 034015 (2019), 1811.07486.

\bibitem{Kasmaei:2019ofu}
B.~S. Kasmaei and M.~Strickland,
\newblock Phys. Rev. D {\bf 102}, 014037 (2020), 1911.03370.

\bibitem{Floerchinger:2021xhb}
S.~Floerchinger, C.~Gebhardt, and K.~Reygers,
\newblock (2021), 2112.12497.

\bibitem{JETSCAPE:2020shq}
JETSCAPE, D.~Everett {\em et~al.},
\newblock Phys. Rev. Lett. {\bf 126}, 242301 (2021), 2010.03928.

\bibitem{Citron:2018lsq}
Z.~Citron {\em et~al.},
\newblock CERN Yellow Rep. Monogr. {\bf 7}, 1159 (2019).

\bibitem{Gale:2021emg}
C.~Gale, J.-F. Paquet, B.~Schenke, and C.~Shen,
\newblock Phys. Rev. C {\bf 105}, 014909 (2022), 2106.11216.

\bibitem{Schafer:2021slz}
A.~Sch\"afer, O.~Garcia-Montero, J.-F. Paquet, H.~Elfner, and C.~Gale,
\newblock Phys. Rev. C {\bf 105}, 044910 (2022), 2111.13603.

\bibitem{Yazdi:2022cuk}
R.~M. Yazdi, S.~Shi, C.~Gale, and S.~Jeon,
\newblock (2022), 2207.12513.

\bibitem{ALICE:2015xmh}
ALICE, J.~Adam {\em et~al.},
\newblock Phys. Lett. B {\bf 754}, 235 (2016), 1509.07324.

\bibitem{Vujanovic:2020wuk}
JETSCAPE, G.~Vujanovic {\em et~al.},
\newblock {Multi-stage evolution of heavy quarks in the quark-gluon plasma},
\newblock in {\em {28th International Conference on Ultrarelativistic
  Nucleus-Nucleus Collisions}}, 2020, 2002.06643.

\bibitem{Fan:2020afj}
JETSCAPE, W.~Fan and G.~Vujanovic,
\newblock PoS {\bf HardProbes2020}, 067 (2021), 2009.04946.

\bibitem{Endres:2015egk}
S.~Endres, H.~van Hees, and M.~Bleicher,
\newblock Phys. Rev. C {\bf 93}, 054901 (2016).

\bibitem{Endres:2016tkg}
S.~Endres, H.~van Hees, and M.~Bleicher,
\newblock Phys. Rev. C {\bf 94}, 024912 (2016).

\bibitem{Staudenmaier:2017vtq}
J.~Staudenmaier, J.~Weil, V.~Steinberg, S.~Endres, and H.~Petersen,
\newblock Phys. Rev. {\bf C98}, 054908 (2018).

\bibitem{Hirayama:2022rur}
R.~Hirayama, J.~Staudenmaier, and H.~Elfner,
\newblock (2022), 2206.15166.

\bibitem{Seck:2020qbx}
F.~Seck {\em et~al.},
\newblock (2020), 2010.04614.

\end{thebibliography}

\end{document}